\documentclass[10pt, reqno]{amsart}

\usepackage{amssymb} 
\usepackage[english, spanish]{babel}
\usepackage[utf8]{inputenc}
\usepackage{cancel}
\numberwithin{equation}{section}

\newenvironment{poliabstract}[1]
  {\renewcommand{\abstractname}{#1}\begin{abstract}}
  {\end{abstract}}

\newtheorem{Theorem}{\sc Teorema}[section]

\newtheorem{Remark}{\sc Observación}[section]

\title[Una Interpretación Física de la Constante de Planck]{Una Interpretación Física de la Constante de Planck}

\author[W. P. \'Alvarez-Samaniego, B. \'Alvarez-Samaniego, D. Moya-Álvarez]{}

\date{December 27, 2012}

\email{alvarezwilson@hotmail.com} 
\email{balvarez@uce.edu.ec}
\email{douglas.moya@epn.edu.ec} 

\begin{document}
\maketitle 

\centerline{\scshape Wilson P. \'Alvarez-Samaniego}
\smallskip
{\footnotesize
 \centerline{N\'ucleo de Investigadores Cient\'{\i}ficos}
   \centerline{Facultad de Ingenier\'{\i}a, Ciencias F\'{\i}sicas y Matem\'atica}
   \centerline{Universidad Central del Ecuador (UCE)}
   \centerline{Quito, Ecuador}}

\vspace{0.5cm}

\centerline{\scshape Borys \'Alvarez-Samaniego}
\smallskip
{\footnotesize
 \centerline{N\'ucleo de Investigadores Cient\'{\i}ficos}
   \centerline{Facultad de Ingenier\'{\i}a, Ciencias F\'{\i}sicas y Matem\'atica}
   \centerline{Universidad Central del Ecuador (UCE)}
   \centerline{Quito, Ecuador}}

\vspace{0.5cm}

\centerline{\scshape Douglas Moya-Álvarez}
\smallskip
{\footnotesize
 \centerline{Departamento de Física}
   \centerline{Facultad de Ciencias}
   \centerline{Escuela Politécnica Nacional (EPN)}
   \centerline{Ladrón de Guevara, S/N}
   \centerline{Quito, Ecuador}}

\selectlanguage{english}
\begin{poliabstract}{Abstract}
According to the commonly accepted interpretation of Quantum Mechanics, it is only possible to talk  
about the existence of elementary particles when they are detected by an experiment or 
by a classical measurement device.  This has led to distortions with regard to the objective 
existence of reality, since it would be necessary an observer to materialize the elementary 
particle.  This issue is solved when the elementary particle is placed in permanent interaction 
with the rest of the universe, so its existence would not depend on any intelligent observer and 
thus the objective range of elementary particles would be recovered.  This article explores the 
interaction between elementary particles and a background action field of stochastic character and 
it is also determined that the standard deviation of the particle-background field interaction 
is essentially the Planck constant, since the expressions of the energy according to the Planck 
postulate and the momentum according to the Louis de Broglie thesis follow in a natural way.  In addition, 
the expression for the wave function of a free particle is obtained from the solution of the integral 
equation that relates the momentum and the energy with the wave vector and the angular frequency 
of a Gaussian distribution.  Also, the Schr\"odinger equation is deduced from the expression 
for the wave function of an elementary particle in a potential field. Finally, the uncertainty 
principle is obtained.  The theory presented in this article rescues the objective nature of Quantum 
Mechanics which implies that elementary particles exist independently of observers since the classical measurement 
device in this theory, introduced by the authors, is the own universe.
\end{poliabstract}

\selectlanguage{spanish}

\begin{poliabstract}{Resumen}
De acuerdo a la interpretación comunmente aceptada de la Mecánica Cuántica solo 
se puede hablar de la existencia de las partículas elementales cuando ellas son 
detectadas en un experimento o con un aparato clásico de medida.  Esto ha conllevado 
a tergiversaciones respecto de la existencia objetiva de la realidad, puesto que 
se requeriría un observador para que se materialice la partícula elemental.  Este 
problema se resuelve cuando se conceptúa a la partícula elemental en interacción 
permanente con el resto del Universo de modo que así su existencia no dependería de 
ningún observador inteligente y se rescataría el rango objetivo de las partículas 
elementales.  En este artículo se estudia la interacción entre las partículas 
elementales y un campo de acción de fondo de carácter estocástico y se determina  
que la desviación estándar de la interacción partícula-campo de fondo es esencialmente la 
constante de Planck, pues aparecen de manera natural las expresiones de la 
energía según el Postulado de Planck y de la cantidad de movimiento de acuerdo a la 
tesis de Louis de Broglie.  Además, se obtiene de forma natural la expresión 
de la función de onda de una partícula libre a partir de la resolución de la 
ecuación integral que liga la cantidad de movimiento y la energía con el vector 
de onda y la frecuencia angular de una distribución gaussiana. También, se deduce  
la ecuación de Schr\"odinger como consecuencia de la expresión de la función de onda 
de una partícula elemental en un campo potencial.  Finalmente, se obtiene el Principio de 
Incertidumbre.  La teoría presentada en este artículo rescata el 
carácter objetivo de la Mecánica Cuántica por el cual las partículas elementales existen  
independientemente de los observadores ya que el aparato de medida clásico en esta teoría, 
introducida por los autores, es el propio Universo.  
\end{poliabstract}

\vspace{1cm}

\maketitle

\section{Introducción}
Sir Isaac Newton (1642-1727) cuando concibió el principio de inercia imaginó la
existencia de un objeto masivo que mantenía su estado en movimiento rectilíneo 
uniforme.  De hecho, Newton suponía que los objetos físicos pueden existir por 
sí mismos, de manera que podemos abstraer su existencia del resto del 
Universo sin alterar las cualidades dinámicas del mencionado objeto.  Este punto 
de vista concordaba con la Filosofía Racionalista de René Descartes (1596-1650) en 
la que se plantea por primera vez en la historia de Occidente la existencia 
de un ser humano independiente del entorno social y por lo tanto la afirmación 
del individuo \textit{en sí y de por sí}, el cual es uno de los principios fundamentales 
sobre los que se levantaría la futura ideología burguesa.  Este elemento ideológico 
penetró en la Ciencia haciendo que esta solamente sea exitosa en el mundo macroscópico, 
donde el experimentador puede controlar las operaciones de medida y sus errores 
haciéndolos tan pequeños como él desee, lo que equivale a considerar el límite clásico 
de la constante de Planck igual a cero. 

Sin embargo, en el mundo microscópico donde existen objetos tan pequeños como son 
las moléculas, átomos, partículas elementales, cuyas dimensiones tienen órdenes menores o 
iguales a $10^{-10}$ m,  esas partículas están inmersas 
en un océano de interacciones cuyo origen está en la propia existencia de 
cada una de ellas extendidas a todo el Universo.  Este océano de interacciones 
fluctúa estocásticamente haciendo imposible determinar simultáneamente 
con precisión la posición y velocidad de las partículas elementales.  En 
este trabajo se demuestra que la función de densidad de la acción a la que está 
sometida la partícula es una gaussiana, lo cual es una consecuencia 
del Teorema Central del Límite.  Expresando esta función en el dominio de la 
frecuencia se encuentra otra distribución gaussiana cuyo exponente es nuevamente 
una función cuadrática de una combinación lineal de las variables espacial y temporal, 
y donde los coeficientes de la variable espacial y de la variable temporal en el 
exponente son el módulo del vector de onda $k$ y la frecuencia angular 
$\omega=\omega(\vec{k})$ de la onda respectivamente.  De ahí se deduce que el 
módulo de la cantidad de movimiento $P$ es proporcional a la desviación estándar 
$\sigma$ del campo de acción de fondo por el módulo del vector de onda, y 
la energía $E$ de la partícula es también proporcional a la frecuencia angular por 
la desviación estándar del campo de acción de fondo.  

Se demuestra además que el núcleo de la ecuación integral que permite transformar 
la representación en el dominio de la frecuencia a la representación en el dominio del 
espacio-tiempo comunes es el producto de dos deltas de Dirac cuyos argumentos son 
$\frac{P}{\hbar}-k$ y $\frac{E}{\hbar} - \omega$ respectivamente.  Expresando estas dos 
distribuciones mediante una superposición continua de ondas planas se concluye que la 
partícula puede ser representada por una función compleja de la forma
\begin{equation}
  \psi(x,t) = \frac{1}{\sqrt{2 \pi}} e^{i \big( \frac{Px-Et}{\hbar} \big)}.
\end{equation}
A partir de la última expresión y usando la relación no relativista de la energía y el hecho 
de que la función de onda del electrón en un potencial, $U(\vec{r})$, se puede expresar como 
una superposición de ondas planas, se procede a deducir en la Sección \ref{sec:sch} 
la ecuación de Schr\"odinger de la partícula elemental.  

En la Sección \ref{sec:inc} se demuestra el Principio de Incertidumbre desde la 
función de densidad de probabilidad de la  acción del campo de fondo de una partícula libre.  Además, se 
obtiene la constante de proporcionalidad entre la desviación estándar y la constante de 
Planck para la Mecánica Cuántica tradicional.  Finalmente, se encuentra la ecuación correspondiente 
a la energía del punto cero del sistema partícula libre-campo de acción de 
fondo (ver ecuación (\ref{eq:energia})).

\section{Función de Distribución de la Acción e Interpretación Física de 
la Constante de Planck}  \label{sec:cte}

Vamos a comenzar deduciendo la densidad de distribuci\'{o}n gaussiana de la acción del  
campo de fondo sobre la partícula.  Para ello, se inicia obteniendo 
la función de densidad que se presenta en un conjunto repetido de mediciones.  De esta 
forma, se determina el valor real de la medición como el promedio de los valores 
experimentales obtenidos y se consigue además la dispersión estadística de datos 
de este proceso.

A continuación, se enuncian los supuestos para la obtención de la densidad de 
distribución gaussiana: 
\begin{enumerate}
\item Peque\~nos errores son más probables que grandes errores.
\item Para un valor real $p$ dado, las dispersiones estadísticas en $\pm \varepsilon$, 
tienen igual probabilidad. 
\item En la presencia de varias observaciones sobre la misma cantidad, el valor 
m\'{a}s probable de esa cantidad es el promedio de las observaciones.
\end{enumerate}

Carl Friedrich Gauss (1777-1855) se refiri\'{o} a dicho proceso como 
``el problema más importante de las Matem\'{a}ticas en la Filosof\'{i}a Natural''.

Ahora, se procede a la deducción de la densidad de distribución gaussiana asociada a 
la acción de la partícula.  Sea $p \in \mathbb{R}$ el verdadero valor (pero desconocido) de 
la medida de la cantidad física.  Se efectúan $n \in \mathbb{N}$ observaciones independientes 
del experimento asociado a la medida de la cantidad física $p$, dichas observaciones dan 
como resultado las medidas $M_{1}, M_{2}, \hdots, M_{n}$.  Sea además $\phi$ la funci\'{o}n de 
densidad de probabilidad del error aleatorio.  Se supone que la funci\'{o}n $\phi$ es diferenciable, 
y que $\phi(x) \neq 0$, para todo $x \in \mathbb{R}$. \\
La suposici\'{o}n 1 anterior implica que $\phi$ tiene un valor m\'{a}ximo en $x=0$, 
mientras que la afirmaci\'{o}n 2 implica que $\phi (x)=\phi (-x)$, para todo $x \in \mathbb{R}$.\\
Se define la función $f: \mathbb{R} \mapsto \mathbb{R}$ por
\begin{equation*}
f(x):=\frac{\phi{}'(x)}{\phi (x)}, \text{ para todo } x \in \mathbb{R}.
\end{equation*}
Entonces,
\begin{equation*}
f(-x)=-f(x),  \text{ para todo } x \in \mathbb{R}.
\end{equation*}
Note que $X_i:= M_{i}-p$ denota la variable aleatoria asociada al error de la i-\'{e}sima 
medida. Ya que estas medidas (y errores) se asumen estoc\'{a}sticamente independientes, se sigue que
\begin{equation*}
\Omega_n :=\phi (M_{1}-p)\phi (M_{2}-p)\hdots \phi (M_{n}-p)=\prod_{i=1}^{n} \phi (M_{i}-p) 
\end{equation*}
es la densidad conjunta asociada a los $n$ errores.  Por otro lado, de la afirmaci\'{o}n 3 se tiene 
que
\begin{equation*}
\bar{M}_n:=\dfrac{M_{1}+M_{2}+....+M_{n}}{n}
\end{equation*} 
es el estimador veros\'{i}mil de $p$. En otras palabras, dadas las medidas 
$M_{1}, M_{2}, \hdots, M_{n}$, al escoger $p=\bar{M}_n$, se maximiza el valor de $\Omega_n$.  \\
A continuación, se evalúa el valor de la derivada de $\Omega_n$ en el punto $p=\bar{M}_n$.
\begin{align*}
&0=\frac{d\Omega_n }{dp}\Big |_{p=\bar{M}_n}  \\
&=- \sum_{i=1}^n \phi^{\prime}(M_{i}-\bar{M}_n)\prod_{j\not\neq i}  \phi(M_{j}-\bar{M}_n)  \\ 
&=- \sum_{i=1}^{n} \dfrac{\phi^{\prime}(M_{i}-\bar{M}_n)}{\phi(M_{i}-\bar{M}_n)}  \prod_{k=1}^n  
\phi(M_{k}-\bar{M}_n)\\
&=- \sum_{i=1}^{n} f(M_{i}-\bar{M}_n) \; \Omega_n \\
& =-\Omega_n \; \sum_{i=1}^n f(M_{i}-\bar{M}_n).
\end{align*}
Entonces,
\begin{equation} \label{eq:1}
\sum_{i=1}^n f(M_{i}-\bar{M}_n)=0.
\end{equation}
Se consideran ahora $M$ y $N$ dos variables aleatorias arbitrarias que podrían representar 
diversas magnitudes físicas tales como longitud, energía u otras.  En vista que las medidas dadas 
por las variables aleatorias $M_{i}$, $i= 1, \hdots, n$, pueden tomar valores arbitrarios, se 
toma a continuación
\begin{equation}  \label{eq:2}
M_{1}=M,\ \ \ M_{2}=M_{3}=\cdots =M_{n}=M-nN.
\end{equation}
Para tal conjunto de medidas, se obtiene por lo tanto que
\begin{equation*}
\bar{M}_n=M-(n-1)N.
\end{equation*}
Luego, de la ecuación (\ref{eq:1}) y considerando (\ref{eq:2}), se tiene que
\begin{equation*}
 f(M-[ M-(n-1)N])+(n-1)f(M-nN-(M-(n-1)N))=0.
\end{equation*}
Por lo tanto,
\begin{equation*}
  f[(n-1)N]=(n-1)f(N).
\end{equation*}
De la paridad y de la continuidad de $f$, se sigue que existe $k \in \mathbb{R}$ tal que 
$f(x)=kx$, para todo $x \in \mathbb{R}$.  Luego, 
\begin{align*}
  &f(\lambda x)=\lambda k \;x =\lambda \; f(x), \text{ para todo } x \in \mathbb{R}.    
\end{align*}
Entonces, 
\begin{equation*}
  \frac{\phi'(x)}{\phi (x)}= kx, \text{ para todo } x \in \mathbb{R}.     
\end{equation*}
De donde se sigue que 
\begin{equation*} 
  \phi(x) = Ae^{\frac{kx^{2}}{2}}, \text{ para todo } x \in \mathbb{R}.  
\end{equation*}
Definamos ahora, 
\begin{equation*}
 k=-\dfrac{1}{\sigma^{2}}.
\end{equation*}
Así,
\begin{equation*}  
\phi (x)=Ae^{-\frac{x^{2}}{2\sigma ^{2}}}, \text{ para todo } x \in \mathbb{R}.
\end{equation*}
De la expresión anterior y suponiendo que $\int_{\mathbb{R}}\phi (x)dx=1$, se observa que 
la constante $A$ está dada por
\begin{equation*}
  A=\frac{1}{\sqrt{2\pi}\sigma}.
\end{equation*}	
En consecuencia, 
\begin{equation}  \label{eq:3}
\phi (x)= \frac{1}{\sqrt{2\pi}\sigma} e^{-\frac{x^{2}}{2\sigma ^{2}}}, 
\text{ para todo } x \in \mathbb{R}.
\end{equation}
Luego, la función de densidad de distribución de la acción del campo de 
fondo viene dada por:
\begin{align*} 
  \phi(S)=\frac{1}{\sqrt{2\pi} \sigma}e^{-\frac{S^{2}}{2\sigma ^{2}}},
\end{align*}
donde $S=Px-Et$, aquí $P$ corresponde al momentum de la partícula y $E$ denota su energía.  

A continuación, se presenta una interpretación de la desviación estándar $\sigma$ en 
función de la constante de Planck $\hbar$ (ver Observación \ref{Obs1}) abajo.  Usando una 
ecuación integral para representar al término $\sqrt{2\pi} \sigma \; \phi(S)$, se puede 
escribir 
\begin{align}   \label{eq:4}	
  \exp(-\frac{S^{2}}{2\sigma ^{2}})=\iint\limits_{\mathbb{R}^2} \ B(P,E,k,\omega)C(kx-\omega t)dkd\omega, 
\end{align}
donde  en el lado izquierdo de la última ecuación aparece esencialmente la función de densidad de distribución 
de la acción S en el espacio-tiempo,  $C(kx-\omega t)$ corresponde a la representación de esa distribución 
en el espacio de Fourier y $B(P,E,k,\omega)$ corresponde al Jacobiano de la transformación entre 
esos dos espacios.  Vale mencionar que este tratamiento fue introducido por A. Einstein en uno de 
sus tres famosos artículos publicados en 1905 para describir el movimiento Browniano y se 
lo considera como el inicio de la teoría de Procesos Estocásticos.   Sea
\begin{align*}
  \theta := kx-\omega t.
\end{align*}
Derivando (\ref{eq:4}) respecto a $x$, y usando formalmente el Teorema de Convergencia Dominada 
(Teorema \ref{tcd}) con las hipótesis de regularidad e integrabilidad apropiadas, se obtiene
\begin{equation*}
  -\frac{S}{ \sigma^{2} }\frac{\partial S}{\partial x}\exp(-\frac{S^{2}}{2 \sigma^{2} })
  =\iint\limits_{\mathbb{R}^2} B(P,E,k,\omega)   C'(kx-\omega t) \frac{\partial \theta }{\partial x }dkd\omega.
\end{equation*}
Derivando ahora (\ref{eq:4}) respecto a la varible $t$, y usando formalmente el Teorema de 
Convergencia Dominada, se ve que
\begin{equation*}
  -\frac{S}{ \sigma^{2} }\frac{\partial S}{\partial t}\exp(-\frac{S^{2}}{2 \sigma^{2} })
	=\iint\limits_{\mathbb{R}^2} B(P,E,k,\omega)  C'(kx-\omega t) \frac{\partial \theta }{\partial t }dkd\omega.
\end{equation*}
Luego,
\begin{align*}
  &-\frac{S}{ \sigma^{2} }\frac{\partial S}{\partial x} \iint\limits_{\mathbb{R}^2} 
	B(P,E,k,\omega) C(\theta )dkd\omega
	=\iint\limits_{\mathbb{R}^2} B(P,E,k,\omega)  C'(\theta) 
	\frac{\partial \theta }{\partial x}dkd\omega,\\
  &-\frac{S}{ \sigma^{2} }\frac{\partial S}{\partial t} 
	\iint\limits_{\mathbb{R}^2} B(P,E,k,\omega) C(\theta )dkd\omega
	=\iint\limits_{\mathbb{R}^2} B(P,E,k,\omega)  C'(\theta) 
	\frac{\partial \theta }{  \partial t }dkd\omega.
\end{align*}
As\'{i},
\begin{align*}
  &\iint\limits_{\mathbb{R}^2} B(P,E,k,\omega) 
	\left [ \frac{S}{\sigma ^{2}}\frac{\partial S}{\partial x} C(\theta)
  +C'(\theta) \frac{\partial \theta }{\partial x }\right ]dkd\omega=0,\\
  &\iint\limits_{\mathbb{R}^2} B(P,E,k,\omega) 
	\left [ \frac{S}{\sigma ^{2}}\frac{\partial S}{\partial t} C(\theta)
  +C'(\theta) \frac{\partial \theta }{\partial t }\right ]dkd\omega=0.
\end{align*}
Ahora, se considera
\begin{equation}  \label{eq:5}
  \frac{S}{\sigma ^{2}} \frac{\partial S}{\partial x} C(\theta)
	+ C'(\theta) \frac{\partial \theta }{\partial x} =0, 
\end{equation}	
y
\begin{equation}  \label{eq:6}
	\frac{S}{\sigma ^{2}}\frac{\partial S}{\partial t} C(\theta)
	+ C'(\theta) \frac{\partial \theta }{\partial t}=0. 
\end{equation}
Multiplicando la ecuación (\ref{eq:5}) por $dx$ y la ecuación (\ref{eq:6}) por $dt$ y sumando, 
se consigue  
\begin{equation*}
\left [ C'(\theta)\frac{\partial \theta }{\partial x}dx
+C'(\theta)\frac{\partial \theta }{\partial t}dt 
+C(\theta) \Big( \frac{S}{\sigma ^{2}}\frac{\partial S}{\partial x}dx
+\frac{S}{\sigma ^{2}}\frac{\partial S}{\partial t}dt \Big) \right ]=0.
\end{equation*}
Luego,
\begin{align*}
 dC(\theta) + C(\theta) d\Big( \frac{S^2}{2 \sigma^2} \Big) = 0.
\end{align*}
Entonces,
\begin{align*}
  \frac{dC(\theta)}{C(\theta)}  + d\Big( \frac{S^2}{2 \sigma^2} \Big)   = 0.
\end{align*}
De donde, integrando se obtiene que
\begin{align*}
  \ln (C(\theta)) + \frac{S^2}{2 \sigma^2}  = \alpha,
\end{align*}
donde $\alpha$ es una constante.  Luego,
\begin{align*}
  C(\theta)  =  e^\alpha \; e^{-\frac{S^{2}}{2\sigma ^{2}}}.
\end{align*}
Así, se tiene que 
\begin{align}  \label{eq:7}
  C(kx-\omega t)  =  e^\alpha e^{-\frac{(Px-Et)^{2}}{2\sigma ^{2}}}, 
	\text{ para todo } x \in \mathbb{R}, \; \; t \in \mathbb{R}.
\end{align}
Tomando $t=0$ en la ecuación (\ref{eq:7}), se sigue que
\begin{align*}
  C(y) = e^\alpha e^{-\frac12 (\frac{P}{\sigma k})^2 y^2}, 
	\text{ para todo } y \in \mathbb{R}.
\end{align*}
Por otro lado, escogiendo $x=0$ en (\ref{eq:7}), se ve que
\begin{align*}
  C(y) = e^\alpha e^{-\frac12 (\frac{E}{\sigma \omega})^2 y^2}, 
	\text{ para todo } y \in \mathbb{R}.
\end{align*}
De las dos expresiones anteriores se obtiene la siguiente relación:
\begin{align} \label{eq:8}
        \frac{P}{k}   =  \frac{E}{\omega}.
\end{align}
\begin{Remark}  \label{Obs1}
Sin embargo, se conoce de la Mecánica Cuántica habitual que $\frac{P}{k} = \frac{E}{\omega} = \hbar$. 
Además, de (\ref{eq:7}) se sigue que $\frac{P}{\sigma}$ es proporcional a $k$, de lo cual se tiene que  
$\sigma$ es proporcional a $\hbar$ (existe $\beta>0$ tal que $\hbar = \beta \sigma$ ).  De aquí en adelante, 
sin pérdida de generalidad, se escoge $\alpha = 0$ en (\ref{eq:7}).
\end{Remark}

Finalmente, se procede a construir la función de onda planar asociada a la 
partícula libre.  Usando la ecuación (\ref{eq:4}) y la constante $\beta>0$ mencionada  
en la  Observación \ref{Obs1}, se tiene que
\begin{eqnarray*}
  e^{-\frac{(Px-Et)^{2}}{2\sigma ^{2}}}
  &=& \iint\limits_{\mathbb{R}^2} \ B(P,E,k,\omega)C(kx-\omega t)dkd\omega  \\
  &=& \iint\limits_{\mathbb{R}^2} B(P,E,k,\omega) e^{-\frac{\beta^2 (kx-\omega t)^2}{2}}dkdw.
\end{eqnarray*}	
Tomando por ejemplo $B(P,E,k,\omega) = \delta (\frac{P}{\beta \sigma}-k)\delta (\frac{E}{\beta \sigma }-\omega)$ se ve que	
\begin{equation*}	
  e^{-\frac{(Px-Et)^{2}}{2\sigma ^{2}}} = 
   \iint\limits_{\mathbb{R}^2} \delta \Big(\frac{P}{\beta \sigma }-k\Big)
	 \delta \Big(\frac{E}{\beta \sigma }-\omega\Big)
	 e^{-\frac{\beta^2 (kx-\omega t)^{2}}{2}}dkd\omega.
\end{equation*}
Usando la {\it{transformada de Fourier}} $\mathcal{F}$, también denotada por \; $\hat{}$ \;, se tiene que
\begin{equation*}
  \widehat{\frac{e^{ia\cdot}}{\sqrt{2\pi}}} = \delta_a, \; \; \text{ para todo } a \in \mathbb{R},
\end{equation*}
donde $\delta_a$, para $a \in \mathbb{R}$, representa la \textit{distribución temperada} definida por 
\begin{equation*}
   \delta_a (\varphi) = \varphi(a), \; \; \text{ para } \varphi \in \mathcal{S}(\mathbb{R}), 
\end{equation*}
aquí $\mathcal{S}(\mathbb{R})$ representa el {\it{espacio de Schwartz}} de las funciones rápidamente 
decrecientes  en $\mathbb{R}$.  Luego, se sigue (usando la notación usual en Física) que
\begin{equation*}
 \delta \Big(\frac{P}{\hbar} -k \Big) = \frac{1}{\sqrt{2\pi}}
 \int_{-\infty}^{+\infty} e^{i\big(\frac{P}{\hbar} -k\big)x} dx 
\end{equation*}
y 
\begin{equation*}
 \delta \Big(\frac{E}{\hbar} -\omega \Big) = \frac{1}{\sqrt{2\pi}}
 \int_{-\infty}^{+\infty} e^{-i\big(\frac{E}{\hbar} -\omega\big)t} dt.
\end{equation*}
Entonces, siguiendo con la notación usual en Mecánica Cuántica, se ve que 
\begin{equation*}
  \delta \Big(\frac{P}{\hbar} -k \Big)  \delta \Big(\frac{E}{\hbar} -\omega \Big)
	= \frac{1}{2\pi}  \iint\limits_{\mathbb{R}^2}  
	  e^{i \frac{(Px-Et)}{\hbar}}  e^{-i(kx-\omega t)}  dxdt.
\end{equation*}
De donde, usando la condición de normalización de Dirac, se observa la representación espacial 
de la {\it{función de onda de la partícula libre}} dada por 
\begin{equation*}
  \psi(x,t) = \frac{1}{\sqrt{2\pi}} e^{i\big(\frac{Px-Et}{\hbar}\big)}.
\end{equation*}

\section{Deducción de la Ecuación de Schr\"odinger} \label{sec:sch}
Si un electr\'{o}n est\'{a} sometido a una energ\'{i}a potencial, $U(\vec{r})$, este forma el 
paquete de ondas
\begin{equation*}
  \psi (\vec{r},t)=\int D(\vec{k})e^{i(\vec{k} \cdot \vec{r}-\omega(\vec{k})t)}d^{3} \vec{k}, 
\end{equation*}
donde $D$ es la representación en el espacio de momentums de la función de onda.  Derivando respecto 
al tiempo la expresión anterior y usando formalmente el Teorema de Convergencia Dominada 
con las hipótesis convenientes, se obtiene que
\begin{equation*}
  \frac{\partial \psi }{\partial t} (\vec{r},t) = 
  \int -i\omega(\vec{k}) D(\vec{k})e^{i(\vec{k} \cdot \vec{r}- \omega(\vec{k}) t)}d^{3} \vec{k}
\end{equation*}
y como $E= \hbar \omega(\vec{k})$, se ve que 
\begin{equation*}
  \frac{\partial \psi }{\partial t} (\vec{r},t) = \int -i\frac{E}{\hbar}
  D(\vec{k})e^{i(\vec{k} \cdot \vec{r}-\omega(\vec{k})t)}d^{3}\vec{k}.
\end{equation*}
Luego, 
\begin{equation*}
  i\hbar\frac{\partial \psi }{\partial t} (\vec{r},t) =
	\int ED(\vec{k})e^{i(\vec{k}\cdot \vec{r}-\omega(\vec{k})t)}d^{3}\vec{k}.
\end{equation*}
En el caso no relativista se conoce que 
\begin{equation*}
E=\dfrac{P^{2}}{2m}+U(\vec{r}).
\end{equation*}
As\'{i} que
\begin{equation*}
  i\hbar \frac{\partial \psi }{\partial t} (\vec{r},t) = 
	\int\frac{P^{2}}{2m}D(\vec{k})e^{i(\vec{k}\cdot \vec{r}-\omega(\vec{k})t)}d^{3}\vec{k} 
	+U(\vec{r}) \; \psi(\vec{r},t).
\end{equation*}
Nuevamente, usando formalmente el Teorema de Convergencia Dominada con las hipótesis oportunas, se 
consigue
\begin{equation*}
 -\bigtriangledown ^{2}\psi (\vec{r},t) = 
 \int k^{2}D(\vec{k}) e^{i(\vec{k} \cdot \vec{r}-\omega(\vec{k})t)}d^{3}\vec{k} 
 =\int \frac{P^{2}}{\hbar^{2}}D(\vec{k})e^{i(\vec{k} \cdot \vec{r}-\omega(\vec{k})t)}d^{3}\vec{k}.
\end{equation*}
De donde se obtiene que 
\begin{equation*}
  -\frac{\hbar^{2}}{2m} \bigtriangledown ^{2}\psi (\vec{r},t) =
	\int \frac{P^{2}}{2m}D(\vec{k})e^{i(\vec{k} \cdot \vec{r}- \omega(\vec{k})t)}d^{3}\vec{k}.
\end{equation*}
En consecuencia
\begin{equation} \label{eq:sh}
  i\hbar\frac{\partial \psi }{\partial t} (\vec{r},t) = 
	-\frac{\hbar^{2}}{2m}\bigtriangledown ^{2}\psi (\vec{r},t) 
	+U(\vec{r})\psi (\vec{r},t), 
\end{equation}
que es la {\it{ecuación de Schr\"odinger}}.

\section{Deducción del Principio de Incertidumbre} \label{sec:inc}
Consideramos una partícula libre cuya función de densidad de probabilidad de la  
acción del campo de fondo viene dada por 
\begin{align*} 
  \phi(S)=\frac{1}{\sqrt{2\pi} \sigma}e^{-\frac{S^{2}}{2\sigma ^{2}}},
\end{align*}
donde 
\begin{align*}
  S = \int_0^t L(t') dt'
\end{align*} 
corresponde a la acción de la partícula,  $L=T-U$ es el Lagrangiano en la 
perspectiva de la Mecánica Clásica no relativista, $T=\frac{P^2}{2m}$ representa la 
energía cinética de la partícula y $U$ es la energía potencial que en el caso de 
la partícula libre puede ser tomada como cero.  Luego, la acción para la partícula 
libre viene dada por
\begin{align*}
  S = \int_0^t \frac{P^2}{2m} dt' = \frac{P^2}{2m} t
	  = \frac{P^2}{2} \frac{x}{mv} = \frac{Px}{2}.
\end{align*}
Por otro lado, la varianza de la acción está expresada por  
\begin{align*}
  \sigma^2 = \overline{S^2} - {\bar{S}}^2,
\end{align*}
donde 
\begin{align*}
   \overline{S^2} = \frac{1}{\sqrt{2\pi} \sigma} \int_{\mathbb{R}} 
	 e^{-\frac{S^{2}}{2\sigma ^{2}}} S^2 dS 
\end{align*}
y
\begin{align*}
   \bar{S} = \frac{1}{\sqrt{2\pi} \sigma} \int_{\mathbb{R}} 
	e^{-\frac{S^{2}}{2\sigma ^{2}}} S dS =0.
\end{align*}
Entonces, 
\begin{align*}
  \sigma^2 = \overline{S^2} = \overline{\Big(\frac{Px}{2}\Big)^2}.
\end{align*}
Por lo tanto, 
\begin{align*}
  \sigma = \frac{1}{2}\sqrt{\overline{(Px)^2}}.
\end{align*}
Como $\sigma$ representa la desviación estándar de la acción del campo de 
fondo de la partícula, entonces cualquier rectángulo en el espacio de 
fase de lados $\Delta x$, $\Delta P$ posee un área mayor o igual que $\sigma$, 
es decir
\begin{align*}
  \Delta x \Delta P \ge \sigma.
\end{align*}
Para coincidir con la Mecánica Cuántica usual se toma $\sqrt{\overline{(Px)^2}} = \hbar$.  
Entonces, se tiene que 
\begin{align} \label{eq:inc}
   \Delta x \Delta P \ge \sigma = \frac{\hbar}{2}.
\end{align}

Se considera ahora el caso de menor incertidumbre, $\Delta x \Delta P = \sigma = \frac{\hbar}{2}$.  
Sea $\varphi$ la función de onda de la partícula libre correspondiente al caso de incertidumbre 
mínima en la representación de coordenadas.  De  la relación 
\begin{equation*}
  \phi(S) dS = |\varphi(x)|^2 dx,
\end{equation*}
la cual determina la probabilidad de que la partícula libre tenga una acción comprendida entre $S$ 
y $S + dS$ y que también es igual a la probabilidad de que la partícula libre en incertidumbre 
mínima esté entre $x$ y $x+dx$, se concluye que
\begin{equation*}
  \varphi(x) = \Big(\frac{P}{\hbar \sqrt{2 \pi}}\Big)^{\frac12} e^{-\frac{P^2x^2}{4 \hbar^2}}.
\end{equation*}
Substituyendo la última expresión en la ecuación de Schr\"odinger se obtiene finalmente que
\begin{equation} \label{eq:energia}
  E = \frac{\hbar \omega}{2} = m c^2, 
\end{equation}
donde $E$ y $m$ son la energía y masa de la partícula libre respectivamente, $c$ es la velocidad 
de la luz y $\omega \equiv \frac{P^2}{2 \hbar m}$ representa la frecuencia natural  de 
oscilación de la partícula en el interior del campo de acción de fondo.  Este último resultado 
será analizado con más detalle en un próximo trabajo.   
\section{Conclusiones}
A continuación algunas consecuencias de lo expuesto en las secciones 
anteriores. 

\begin{itemize}
\item
Se rescata el carácter objetivo de las partículas elementales, puesto que el aparato 
macroscópico de medida constituye el resto del Universo.

\item
Se recupera la naturaleza causal de la Teoría Cuántica, ya que las transiciones 
``espontáneas'' son producto de las perturbaciones del campo de fondo que actúa sobre 
una partícula en estado cuántico ``excitado'' volviendo al estado estacionario de 
menor energía. 

\item
De la Sección \ref{sec:cte}  se deduce que la constante de Planck es ``esencialmente'' 
la desviación estándar del campo de interacciones de fondo con la partícula.

\item
De la ecuación (\ref{eq:8}) y de la Observación \ref{Obs1} se deducen los Postulados 
de Planck y de-Broglie, estableciendo el carácter de la dualidad onda-corpúsculo 
de la Mecánica Cuántica tradicional.

\item
En la Sección \ref{sec:sch} se deduce la ecuación de Schr\"odinger asociada 
a una partícula en un potencial, $U(\vec{r})$, la cual viene dada por la ecuación 
(\ref{eq:sh}).  Esta ecuación se obtuvo suponiendo que la función de onda de dicha 
partícula es una superposición de infinitas ondas planas.  Por otro lado, en la 
Mecánica Cuántica tradicional la ecuación de Schr\"odinger se la postula como 
su principio dinámico (ver \cite{gp} para más detalles).

\item
En la Sección \ref{sec:inc} se deduce, desde la teoría presentada en este artículo,  
el Principio de Incertidumbre.  Además, se observa que para el caso de la Mecánica 
Cuántica habitual, la constante de proporcionalidad mencionada en la Observación  
\ref{Obs1} es $\beta=2$.  Finalmente, se analiza la conducta física de una partícula 
libre en condiciones de incertidumbre mínima, obteniendo la ecuación (\ref{eq:energia})  
que corresponde a la energía del punto cero del sistema partícula-campo de acción 
de fondo.

\end{itemize}

\section{Apéndice}
\begin{Theorem}[Teorema de Convergencia Dominada de Lebesgue \cite{Bartle}]  \label{tcd}
Sea $(\Omega, \mathcal{A}, \mu)$ un espacio de medida.  Sea $(f_n)_{n \in \mathbb{N}}$ 
una sucesión de funciones integrables ($f_n \in L(\Omega, \mathcal{A}, \mu)$, 
para todo $n \in \mathbb{N}$) la cual converge en casi todas partes 
a una función medible real-valuada $f$.  Si existe una función integrable $g$ tal 
que $|f_n| \le g$ para todo $n \in \mathbb{N}$, entonces $f$ es integrable y 
\begin{equation*}
  \int f d\mu = \lim_{n \rightarrow +\infty} \int f_n d\mu.
\end{equation*}
\end{Theorem}

\vspace{2cm}


\begin{thebibliography}{8.}
\bibitem{Bartle} R. G. Bartle, \textit{The Elements of Integration and Lebesgue 
Measure}, Wiley-Interscience; 1 edition (1995). 

\bibitem{Douglas} D. Moya-Álvarez, \textit{El campo de acción.  Una nueva 
interpretación de la Mecánica Cuántica}, editado por la Escuela Politécnica 
Nacional (1994), Quito, Ecuador.

\bibitem{gp} A. Galindo \& P. Pascual, \textit{Quantum Mechanics I}, Springer, 1 edition (1990).


\end{thebibliography}
\end{document}